\documentclass[%
reprint,
amsmath,amssymb,
aps,
]{revtex4-2}
\usepackage{amssymb} 
\usepackage{amsmath}
\usepackage{multirow}
\usepackage{booktabs}

\usepackage{graphicx}
\usepackage{dcolumn}
\usepackage{bm}

\usepackage[version=4]{mhchem}

\begin{document}
	
\preprint{APS/123-QED}
	
\title{\textbf{Anion Engineering Enables Concurrent Optimization of Phonon and Carrier Transport in AMg$_2$X$_2$ Zintl Thermoelectrics}}
\author{Ning Wang$^{1,2,\dag}$, Jincheng Yue$^{3,4,1,\dag,*}$, Hui Zhang$^{1}$, Xingchen Shen$^{2}$, Wenling Ren$^{5}$, Shuyao Lin$^{6,7}$, Yanhui Liu$^{1,*}$, Yue-Wen Fang$^{{3,8},*}$, Tian Cui$^{1,9}$}

\affiliation{$^{1}$ Institute of High-Pressure Physics, School of Physical Science and Technology, Ningbo University, Ningbo 315211, China}

\affiliation{$^2$ MOE Key Laboratory of Materials Physics and Chemistry under Extraordinary Conditions \& Shaanxi Provincial Key Laboratory of Condensed Matter Structure and Properties, School of Physical Science and Technology, Northwestern Polytechnical University, Xi'an, 710072 P. R. China}

\affiliation{$^3$ Centro de Física de Materiales (CFM-MPC), CSIC-UPV/EHU, Manuel de Lardizabal Pasealekua 5, 20018 Donostia/San Sebastián, Spain}

\affiliation{$^{4}$ Fisika Aplikatua Saila, Gipuzkoako Ingeniaritza Eskola, University of the Basque Country (UPV/EHU), Europa Plaza 1, 20018 Donostia/San Sebastián, Spain}

\affiliation{$^5$ Institute of Materials Science, Technical University of Darmstadt, Alarich-Weiss-Strasse 2, Darmstadt, 64287, Germany}

\affiliation{$^6$ Technische Universit\"{a}t Wien, Institute of Materials Science and Technology, Vienna, A-1060, Austria}

\affiliation{$^7$ Link\"{o}ping University, Department of Physics, Chemistry, and Biology (IFM), Link\"{o}ping, SE-58183, Sweden}

\affiliation{$^8$ Instituto de Nanociencia y Materiales de Aragón (INMA), CSIC-Universidad de Zaragoza, Zaragoza,
50009, Spain}

\affiliation{$^9$ State Key Laboratory of High Pressure and Superhard Materials, College of Physics, Jilin University, Changchun, 130012, China}

\date{\today}


\begin{abstract}
The simultaneous suppression of lattice heat transport and enhancement of carrier transport is central to the development of high-performance thermoelectric materials. Here, we investigate the microscopic mechanisms governing phonon and carrier transport in CaAl$_2$Si$_2$-type AMg$_2$X$_2$ (A = Ca, Sr, Ba; X = P, As, Sb, Bi) Zintl compounds using first-principles calculations, machine-learning interatomic potentials, and Boltzmann transport theory. As X changes from P to Bi, the widening phonon band gap restricts three-phonon scattering channels and increases the relative importance of four-phonon processes. In $\mathrm{CaMg_2Bi_2}$ and $\mathrm{BaMg_2Bi_2}$, four-phonon scattering dominates the relaxation of low-frequency heat-carrying modes at 900 K. The Bi-based series also exhibits an anomalous increase in lattice thermal conductivity from Ca to Ba despite the increasing cation mass. Mode-resolved analysis links the lower thermal conductivity of $\mathrm{CaMg_2Bi_2}$ to reduced low-frequency group velocities and stronger anharmonic scattering, with Ca-related mid-frequency vibrations providing additional four-phonon scattering channels. Including both particle-like and coherent contributions, its lattice thermal conductivity reaches $0.23~\mathrm{W\cdot\,m^{-1}\,K^{-1}}$ at 900 K. Meanwhile, the increasing covalent character of the vertical Mg–X bonds is accompanied by enhanced band-edge dispersion and reduced carrier effective masses. These electronic changes, together with conduction-band convergence, favor n-type transport and yield a predicted peak $ZT$ of 3.68 at 900 K in $\mathrm{CaMg_2Bi_2}$. These results elucidate the interplay of phonon spectral structure and higher-order anharmonicity in thermal transport, alongside the distinct electronic response to anion substitution.

\end{abstract}

\maketitle
\section{Introduction}
\label{sec1}
Zintl compounds are a class of polar intermetallic materials with distinctive bonding and structural characteristics, first proposed and developed by Edward Zintl and Wilhelm Klemm in the 1930s \cite{zintl1939intermetallische, klemm1941eduard}. Owing to their diverse chemical compositions and structural flexibility, Zintl compounds provide a rich platform for the design and optimization of thermoelectric materials. Numerous studies have shown that these materials typically exhibit the characteristic "phonon-glass, electron-crystal" (PGEC) behavior, which has attracted considerable attention in thermoelectric research \cite{kauzlarich2007zintl, toberer2010zintl, sarkar2025crystalline}. Since the initial report on the thermoelectric properties of the CaZn$_2$Sb$_2$/YbZn$_2$Sb$_2$ solid solution in 2006 \cite{gascoin2005zintl}, Zintl compounds have garnered considerable interest in thermoelectric research. For instance, the classical Zintl compounds $n$-type Mg$_3$Sb$_2$  achieves a $ZT$ of 2.0 and a single-leg efficiency of 12.6\% by In doping and prolonged sintering that synergistically optimize electron and phonon transport \cite{wang2024high}. Similarly, NaSrSb compound exhibits ultra-low lattice thermal conductivity, achieving an optimal $ZT$ of 2.0 at 900 K \cite{vishwakarma2023ab}. These examples highlight the great potential of Zintl compounds for high-performance thermoelectrics, motivating further exploration of undiscovered members and compositional tuning strategies.

As illustrated in FIG.~\ref{fig1}(a), Zintl compounds can be classified into several stoichiometric types, including 11‑6‑12, 3‑1‑3, 5‑2‑6, 14-1-11, and 1‑2‑2 structures \cite{toberer2008traversing,zevalkink2012thermoelectric,chanakian2015enhanced,hu2015yb14mgsb11,jin2021manipulation}. Among these, the 1-2-2 type Zintl compounds offer multiple degrees of freedom for synergistic anion/cation engineering, as all three crystallographic sites are amenable to substitution. Structurally, 1-2-2 Zintl compounds can be further classified into several representative prototypes, including BaCu$_2$S$_2$, CaAl$_2$Si$_2$, and ThCr$_2$Si$_2$-type structures, among which the CaAl$_2$Si$_2$-type has attracted particular attention in thermoelectric study. The CaAl$_2$Si$_2$-type Zintl compounds exhibit a two-dimensional layered structure consisting of alternately stacked anionic and cationic layers. The anionic framework, formed by a covalently bonded network, provides an efficient pathway for carrier transport, exhibiting "electron-crystal" behavior. Moreover, the interlayer cations interact with the framework via relatively weak ionic bonding, which enhances lattice anharmonicity and significantly suppresses lattice thermal conductivity, giving rise to "phonon-glass" behavior. This unique combination of intrinsically low lattice thermal conductivity and efficient electrical transport makes CaAl$_2$Si$_2$-type compounds promising candidates for high-performance thermoelectric materials \cite{zhang2016designing, zhang2019insights}. Furthermore, the widespread application of thermoelectric technology calls for high-performance thermoelectric materials that are low-cost, environmentally friendly, and non-toxic. Current commercial materials, such as Bi$_2$Te$_3$ or PbTe \cite{chen2023solvothermally, deng2025thermoelectric}, contain large amounts of rare tellurium or toxic element, which hinders their large-scale application. In contrast, Mg-based CaAl$_2$Si$_2$-type Zintl compounds are composed of earth abundant elements, offering low cost and good environmental compatibility, thereby providing a practical foundation for large-scale applications.

In recent years, Mg-based Zintl compounds with CaAl$_2$Si$_2$-type structure have attracted considerable attention in thermoelectrics \cite{li2023wide, zhao2024plasticity}. To further enhance their thermoelectric performance, various tuning strategies have been proposed, including doping, defect engineering, and anion engineering \cite{shuai2016higher, zheng2021defect, wang2025achieving}. For example, in the CaMg$_2$Sb$_2$ system, which suffers from low carrier mobility caused by strong polar optical phonon scattering, a synergistic approach combining isovalent alloying with Cd and Yb together with aliovalent Li doping not only improves carrier mobility but also effectively reduces lattice thermal conductivity, ultimately achieving a $ZT$ of 1.1 at 773 K \cite{guo2025strong}. Furthermore, Wang et al. achieved a conversion efficiency of 9.8\% at 875 K in YbMg$_2$Bi$_{1.98}$Sb$_{0.02}$ by partially substituting Bi with Sb in YbMg$_2$Bi$_2$, where the difference in atomic radius and mass between Bi and Sb introduced lattice distortion \cite{wang2020enhanced}. Although these strategies have led to improved thermoelectric performance, the specific role of anions and the underlying microscopic mechanisms remain insufficiently understood. This motivates us to further investigate the role played by anions in AMg$_2$X$_2$ (A = Ca, Sr, Ba; X = P, As, Sb, Bi) compounds.

In this work, we carry out a comprehensive investigation of the anion-dependent thermoelectric properties of CaAl$_2$Si$_2$-type Zintl compounds AMg$_2$X$_2$ (A = Ca, Sr, Ba; X = P, As, Sb, Bi) by combining first-principles calculations with machine-learning interatomic potentials (MLIPs). The AMg$_2$X$_2$ family exhibits metavalent bonding characteristics and weaker bonding interactions than the structurally analogous Mg$_3$Sb$_2$, providing a favorable chemical basis for suppressing lattice thermal conductivity. Furthermore, the anion-dependent bonding evolution further modifies the lattice vibration and enhances anharmonicity phonon scattering, thereby leading to strongly suppressed lattice thermal conductivity. When the A-site cation is fixed, the contribution of four-phonon scattering becomes increasingly pronounced and gradually dominates the suppression of lattice thermal conductivity as the X-site anion varies from P to Bi. Consequently, CaMg$_2$Bi$_2$ reaches an ultralow lattice thermal conductivity ($\kappa_L$) of $0.23~\mathrm{W\cdot\,m^{-1}\,K^{-1}}$ at 900 K. Moreover, for a fixed A-site cation, changing the X-site anion from P to Bi reduces the electronegativity difference and enhances the covalent character of Mg–X bonding. This bonding evolution leads to a narrowed band gap and reduced carrier effective mass. These electronic structure changes collectively enhance electrical transport, enabling CaMg$_2$Bi$_2$ to achieve a high $n$-type $PF$ of $7.62~\mathrm{mW\cdot\,m^{-1}\,K^{-2}}$ at 900 K. As a result, anion engineering enables the decoupling of thermal and electrical transport. Benefiting from this favorable combination, Bi-based compounds exhibit superior thermoelectric performance, with CaMg$_2$Bi$_2$ achieving a peak $n$-type $ZT$ of 3.68 at 900 K. These findings clarify the coupled role of anion engineering in regulating phonon and carrier transport, and provide a viable strategy for designing high performance Mg-based 1-2-2 Zintl thermoelectric materials.


\section{Computational method}

\subsection{First-principles calculation}
In this work, all density functional theory (DFT) calculations were performed using the projector augmented-wave (PAW) method \cite{blochl1994projector} implemented in the Vienna $ab$ $initio$ simulation package (VASP) \cite{PhysRevB.54.11169,vasp}. The exchange–correlation functional was treated within the generalized gradient approximation (GGA) in the Perdew–Burke–Ernzerhof revised for solids (PBEsol) \cite{PhysRevLett.100.136406}. A plane-wave cutoff energy of 600 eV and a 12×12×12 Monkhorst–Pack $k$-point mesh were employed. The convergence criteria for total energy and atomic forces were set to $10^{-8}$ eV and $10^{-6}$ eV$\cdot$\AA$^{-1}$, respectively. The convex-hull diagrams of all compounds were obtained from the Open Quantum Materials Database (OQMD) \cite{kirklin2015open, saal2013materials}, as shown in FIG. S1. Convex hull analysis indicates that all compounds lie exactly on the convex hull (with a stability of 0 meV$\cdot$atom$^{-1}$), confirming their thermodynamic stability.

\subsection{Machine-learning interatomic potentials}
For machine-learning interatomic potentials, $ab$ $initio$ molecular dynamics (AIMD) \cite{method5} simulations were performed on a 4 × 4 × 2 supercell containing 160 atoms at 20, 50, 100, 200, and 300 K. The simulations were carried out for 3000 time steps, with atomic configurations sampled every 50 steps. Subsequently, static self-consistent calculations were carried out using a 3 × 3 × 3 Monkhorst–Pack $k$-point mesh to obtain training data including atomic configurations with corresponding total energies and atomic forces. A deep potential (DP) \cite{zhang2018deep} model was trained using the DeepMD-kit package \cite{wang2018deepmd} for 1,000,000 steps. During training, the loss function weights were adaptively tuned: the energy and virial weights were increased from 0.02 to 1, while the force weight decreased from 1000 to 1. The accuracy of the trained DP model was validated by comparing the predicted energies and forces with DFT results, as demonstrated in FIGs. S2 and S3 of the Supporting Information. Furthermore, molecular dynamics simulations were performed using the DP model in the canonical (NVT) ensemble at temperatures of 300, 600, and 900 K, with a timestep of 10 fs for a total simulation time of 100 ps as implemented in the FHI-vibes package \cite{knoop2020fhi}. The corresponding energy evolution profiles, shown in FIG. S4 of the Supporting Information, confirm the stability and reliability of the DP model over the investigated temperature range. In addition, moment tensor potential (MTP) \cite{shapeev2016moment} were constructed using the machine-learning interatomic potential (MLIP) package \cite{novikov2021mlip}. The same training dataset as that used for the DP model was employed. The weights for energy, force, and stress in the training process were set to 1, 0.1, and 0.001, respectively. The reliability of the MTP was validated by comparing the predicted energies and forces with DFT results, as illustrated in FIGs. S5 and S6 of the Supporting Information.

\subsection{Phonon properties and thermal transport calculation}
Phonon properties were calculated using the PHONOPY package \cite{method4}. A 4 × 4 × 2 supercell and a 3 × 3 × 3 $k$-point mesh were employed. Second-order force constants were obtained using the finite displacement method \cite{method6}. Third- and fourth-order interatomic force constants were extracted from the trained MTP model using the Thirdorder \cite{method7} and Fourthorder \cite{han2022fourphonon} packages, significantly reducing computational cost. The accuracy of the MTP was verified by comparing phonon dispersion relations with DFT results, as shown in FIG. S7 of the Supporting Information. The lattice thermal conductivity was calculated by solving the Boltzmann transport equation using the ShengBTE package \cite{method7}. Convergence tests on the $q$-point mesh and force-constant cutoff radius were performed to ensure reliability, as demonstrated in FIG. S8 of the Supporting Information. Based on these tests, a $q$-point mesh of 18 × 18 × 18 was adopted for Brillouin zone sampling. The third-order interatomic force constants were computed by considering interactions up to the 16th nearest neighbors, while the fourth-order force constants included interactions up to the 5th nearest neighbors. In addition, an adaptive Gaussian smearing parameter \cite{bjorkman2011adaptive} of 0.1 was employed in the phonon scattering calculations.
\begin{figure*}[htbp]
    \centering
    \includegraphics[width=1\linewidth]{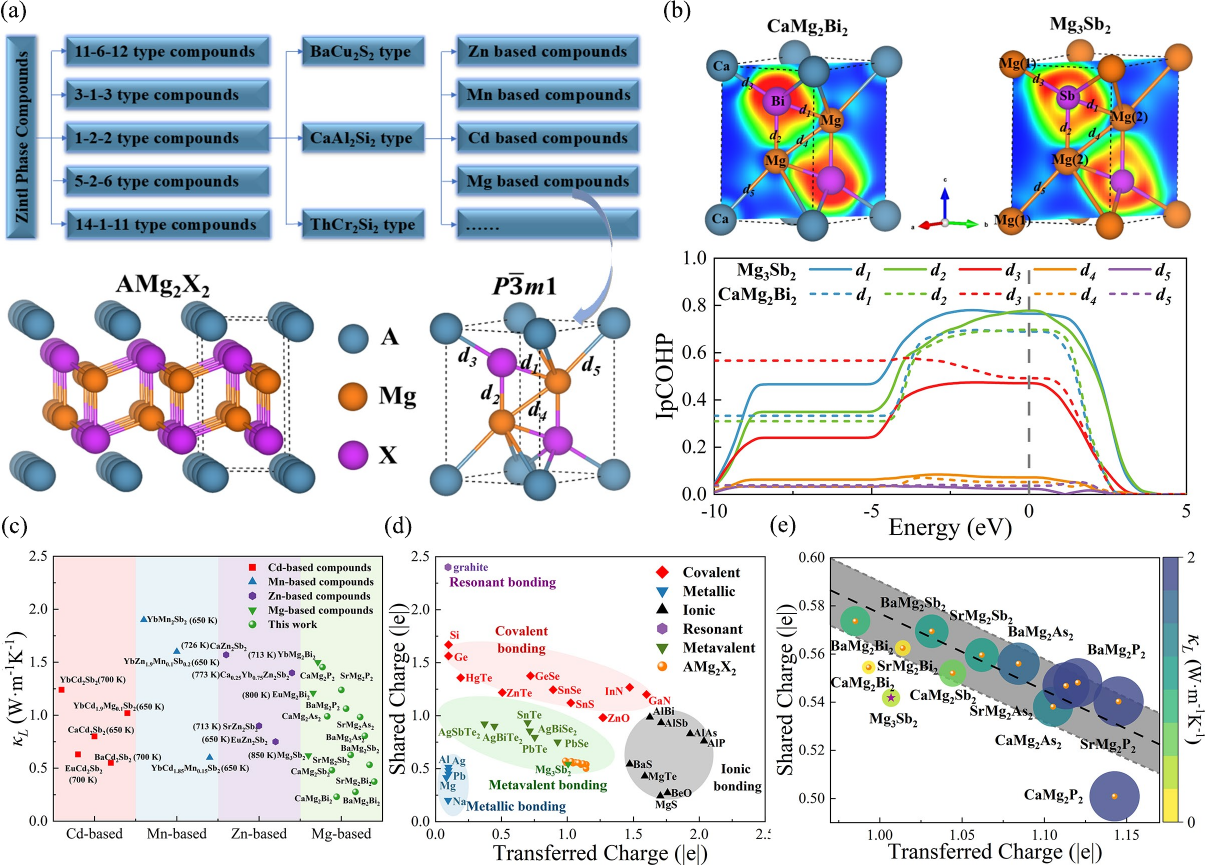}
    \caption{(a) Material screening process for Mg-based CaAl$_2$Si$_2$-type Zintl compounds and crystal structure of AMg$_2$X$_2$ (A = Ca, Sr, Ba; X = P, As, Sb, Bi) compounds. (b) Comparison of the electron localization function (ELF) along the (110) plane and the negative partial integrated crystal orbital Hamilton population (-IpCOHP) for $d$$_1$, $d$$_2$, $d$$_3$, $d$$_4$, and $d$$_5$ bonds in CaMg$_2$Bi$_2$ and Mg$_3$Sb$_2$. (c) Comparison of the lattice thermal conductivity of AMg$_2$X$_2$ compounds at 900 K with those of Cd-, Mn-, and Zn-based CaAl$_2$Si$_2$-type compounds at elevated temperatures, where the data for the representative compounds are taken from Refs \cite{shuai2017recent, peng2018crystal, agne2018heat}. (d) Charge sharing and charge transfer map including metallic, ionic, covalent, and metavalent bonding. The rest of the data in this chart is from Ref \cite{wuttig2018incipient} (e) Lattice thermal conductivity of AMg$_2$X$_2$ at 900 K as a function of transferred charge.}
    \label{fig1}
\end{figure*}

\subsection{Electrical transport properties calculation}
Electronic structure calculations were carried out using the Heyd-Scuseria-Ernzerhof for solids (HSEsol) to obtain accurate band structures \cite{schimka2011improved}. Transport properties were computed using the AMSET package \cite{ganose2021efficient}, which considers acoustic deformation potential scattering (ADP), polar optical phonon scattering (POP), and ionized impurity scattering (IMP). The carrier relaxation time was evaluated using Matthiessen’s rule \cite{Ziman1960}, and the Seebeck coefficient and electrical conductivity were calculated within the Boltzmann transport framework \cite{boltzmann1970weitere}. To ensure well-converged transport properties, a dense interpolated $k$-point mesh was employed. Specifically, an interpolation factor of 50 was adopted, corresponding to an effective $k$-mesh of 73 × 73 × 39.

\section{Results and Discussion}
\par The Zintl compounds AMg$_2$X$_2$ (A = Ca, Sr, Ba; X = P, As, Sb, Bi) belong to the CaAl$_2$Si$_2$-type family which crystallize in space group $P$$\bar{3}$$m$1 (no. 164). As shown by the crystal structure maps in FIG.~\ref{fig1}(a), the structure of AMg$_2$X$_2$ consists of two-dimensional [Mg$_2$X$_2$]$^{2-}$ sheets built from MgX$_4$ tetrahedra that share three of their edges with neighboring tetrahedra. The charge balance is maintained by layers of A$^{2+}$ ions in the $a$-$b$ plane, where each A$^{2+}$ is surrounded by six X atoms as nearest neighbors and six Mg atoms as next-nearest neighbors. Notably, the alternating stacking of anionic and cationic layers along the $c$‑axis forms a layered structure, which is beneficial for blocking phonon movement and limiting heat flow~\cite{wang2024enhanced,li2024layered,yue2024ultralow,yue2025interlayer}. 
 
In FIG.~\ref{fig1}(b), the electron localization function (ELF) and crystal orbital Hamilton population (-IpCOHP) of AMg$_2$X$_2$ are presented, as well as the comparison with Mg$_3$Sb$_2$. Generally, the ELF provides real-space information on electron localization and offers intuitive insight into the nature of chemical bonding. As evidenced in FIG.~\ref{fig1}(b), pronounced charge localization between the anionic and cationic layers of Mg$_3$Sb$_2$ (denoted as the $d$$_3$ bond), indicative of polar covalent character. Although AMg$_2$X$_2$ adopts a similar layered structure, CaMg$_2$Bi$_2$ exhibits a clear absence of electron localization at the corresponding $d$$_3$ bond site, suggesting that the Ca–Bi interaction is dominated by ionic interactions. This bonding behavior is clearly distinct from that of Mg$_3$Sb$_2$.  Meanwhile, the corresponding ‑IpCOHP results demonstrate that both compounds exhibit negligible bonding interactions at the $d$$_4$ and $d$$_5$ sites. Moreover, the $d$$_3$ bond exhibits comparable strength between the two compounds and is markedly weaker than the $d$$_1$ and $d$$_2$ bonds. Such bond softening promotes softer phonon dispersion and expands the phonon scattering phase space, which consequently leads to a significant suppression of $\kappa_L$ \cite{ding2021soft}. Furthermore, compared to Mg$_3$Sb$_2$, the markedly weaker $d$$_1$ and $d$$_2$ bonds in CaMg$_2$Bi$_2$ imply stronger lattice anharmonicity. This increased anharmonicity is expected to more effectively hinder phonon transport, thus resulting in a reduced $\kappa_L$.
\begin{figure*}[htbp!]
    \centering
    \includegraphics[width=1\linewidth]{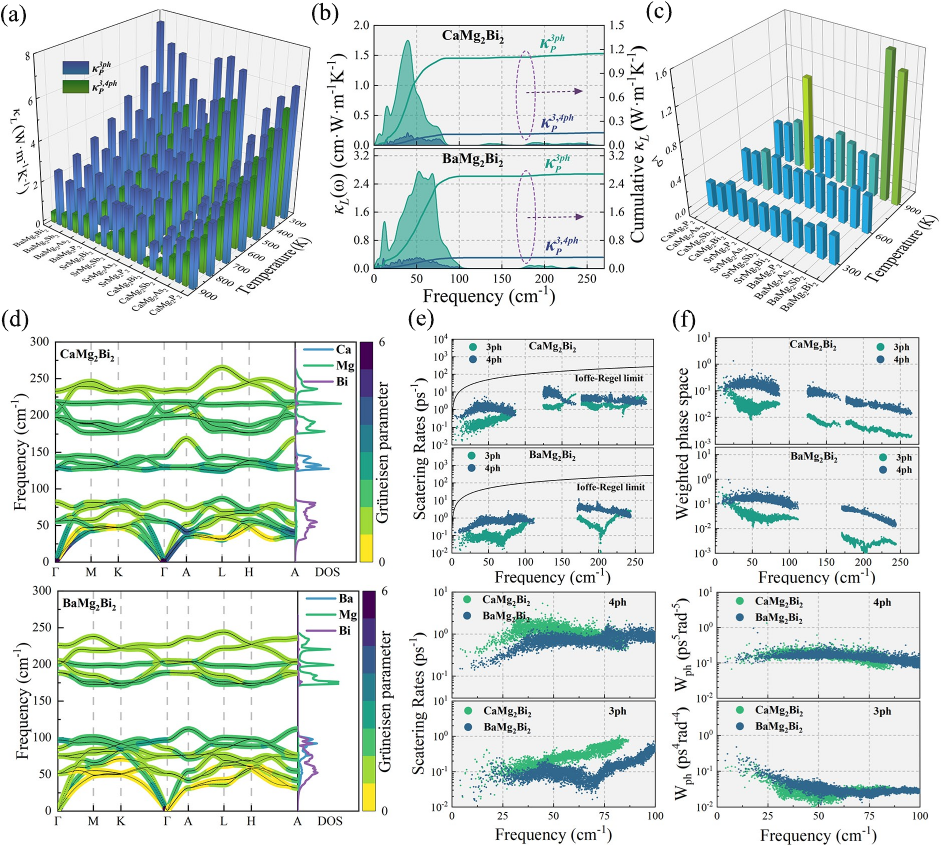}
    \caption{(a) The lattice thermal conductivities of AMg$_2$X$_2$ in the temperature range of 300-900 K. (b) Frequency-dependent cumulative and differential lattice thermal conductivity of CaMg$_2$Bi$_2$ and BaMg$_2$Bi$_2$. (c) Anharmonicity measures of AMg$_2$X$_2$ at 300 K, 600 K, and 900 K. (d) Calculated phonon dispersion of CaMg$_2$Bi$_2$ and BaMg$_2$Bi$_2$. Color smear presents mode Grüneisen parameters, which range from $-$6 to 6. (e) Frequency-dependent three- and four-phonon scattering rates of CaMg$_2$Bi$_2$ and BaMg$_2$Bi$_2$ at 900 K. The upper part illustrates the comparison between three- and four-phonon processes within each material, while the lower section presents a comparison of the three-phonon and four-phonon scattering rates between the two compounds below 100 cm$^{-1}$. (f) Frequency-dependent three- and four-phonon weighted phase spaces of CaMg$_2$Bi$_2$ and BaMg$_2$Bi$_2$ at 900 K. The upper panel compares three- and four-phonon contributions within each material. The lower panel compares the three-phonon and four-phonon-weighted phase spaces between the two compounds below 100 cm$^{-1}$.}
    \label{fig2}
\end{figure*}
 
The $\kappa_L$ of the AMg$_2$X$_2$ compounds was evaluated by considering both particle-like and coherent heat-transport channels, with the total contribution expressed as $\kappa_L = \kappa_P + \kappa_C$, where $\kappa_P$ and $\kappa_C$ represent the particle-like and coherent components, respectively \cite{yue2025diffuson}. The individual contributions are summarized in FIG. S9 of the Supporting Information. As shown in FIG.~\ref{fig1}(c), the AMg$_2$X$_2$ compounds exhibit markedly low $\kappa_L$ at elevated temperatures compared with several representative 1–2–2 Zintl compounds, highlighting their potential for thermoelectric applications. Notably, CaMg$_2$Bi$_2$ reaches an ultralow $\kappa_L$ of $0.23~\mathrm{W\cdot\,m^{-1}\,K^{-1}}$ at 900 K, which is not only lower than those of the benchmark Zintl systems considered here, but also substantially smaller than that of the structurally analogous Mg$_3$Sb$_2$ compound, whose $\kappa_L$ is $0.62~\mathrm{W\cdot\,m^{-1}\,K^{-1}}$ at the same temperature \cite{hu2023breaking}. The pronounced suppression of $\kappa_L$ in CaMg$_2$Bi$_2$ can be attributed to its enhanced lattice anharmonicity, which originates from weakened chemical bonding, as supported by the ELF and $-$IpCOHP analyses discussed above.

Understanding the underlying chemical bonding is essential for rationalizing the thermoelectric performance of AMg$_2$X$_2$ compounds. Here, the net atomic charge and overlap population are used as effective descriptors to identify bonding characteristics and assess bond strength. As shown in FIG.~\ref{fig1}(d), different bonding mechanisms, including covalent, metallic, ionic, and metavalent bonding, can be distinguished in a two-dimensional bonding map defined by the shared and transferred charges. In particular, metavalent bonding occupies a well-defined region characterized by a small transferred charge and a shared charge close to unity, as highlighted in green. Previous studies have suggested that metavalent bonding plays a critical role in enabling the unusual transport properties of high-performance thermoelectric materials \cite{yue2024role,liu2024doping}. Notably, the AMg$_2$X$_2$ compounds are located within the metavalent-bonding region, indicating that this family possesses bonding features favorable for thermoelectric applications.

The correlation between $\kappa_L$ and transferred charge for AMg$_2$X$_2$ compounds at 900 K is further presented in FIG.~\ref{fig1}(e). A general increase in $\kappa_L$ is observed with increasing transferred charge. To clarify the origin of this trend, we systematically compared the $-$IpCOHP values of all investigated compounds, as summarized in FIG. S10 of the Supporting Information. The results show that stronger charge transfer is accompanied by enhanced bond strength, which suppresses lattice anharmonicity and consequently leads to higher $\kappa_L$. Notably, the tilted Mg–X bonds ($d_3$) exhibit a more pronounced dependence on composition. This suggests that the evolution of Mg–X bonding plays a dominant role in determining lattice dynamics and thermal transport behavior. These findings establish a direct connection between chemical bonding, lattice anharmonicity, and heat transport in the AMg$_2$X$_2$ system.

We first examine the $\kappa_P$, which is governed by phonon transport within the Boltzmann transport framework. As shown in FIG.~\ref{fig2}(a), calculations considering only three-phonon scattering substantially overestimate $\kappa_P$, highlighting the essential role of four-phonon processes in suppressing thermal transport. After four-phonon scattering is included, $\kappa_P^{3,4\mathrm{ph}}$ decreases monotonically with increasing X-site atomic mass, accompanied by an increasing relative contribution from four-phonon scattering. This behavior can be attributed to the enlarged phonon band gap in AMg$_2$X$_2$ compounds with heavier X atoms, as shown in FIG. S11, which originates from the reduced vibrational frequencies and enhanced mass contrast. The widened gap restricts the phase space available for three-phonon scattering under energy-conservation constraints, thereby enhancing the relative importance of four-phonon processes.

From classical considerations, materials with similar crystal structures are generally expected to exhibit reduced lattice thermal conductivity with increasing average atomic mass \cite{keyes1959high,slack1973nonmetallic}. However, an anomalous mass-dependent trend is observed for the A-site cation: $\kappa_P^{3,4\mathrm{ph}}$ increases from 0.158 to 0.208 and $0.321~\mathrm{W\cdot\,m^{-1}\,K^{-1}}$ for CaMg$_2$Bi$_2$, SrMg$_2$Bi$_2$, and BaMg$_2$Bi$_2$ at 900 K, respectively. Upon further including the coherent contribution, the total lattice thermal conductivity remains ultralow, with CaMg$_2$Bi$_2$ reaching a minimum $\kappa_L$ of approximately $0.23~\mathrm{W\cdot\,m^{-1}\,K^{-1}}$. Importantly, the anomalous A-site mass dependence persists after the coherent contribution is included, indicating that this behavior is intrinsic to the particle-like phonon transport rather than arising from coherence effects. This suggests that variations in phonon dispersion and anharmonic scattering, rather than atomic mass alone, play a dominant role in determining the thermal transport behavior.

To further elucidate the anomalous A-site mass dependence of thermal transport, we analyzed the frequency-resolved cumulative $\kappa_P$ and differential $\kappa_P$ spectra of CaMg$_2$Bi$_2$ and BaMg$_2$Bi$_2$, as shown in FIG.~\ref{fig2}(b). Including four-phonon scattering further suppresses $\kappa_P$ over the entire frequency range. The dominant contributions to $\kappa_P$ arise from low-frequency phonons, mainly within 9–90 cm$^{-1}$ for BaMg$_2$Bi$_2$ and 14–80 cm$^{-1}$ for CaMg$_2$Bi$_2$, owing to their relatively large group velocities and high phonon occupation numbers. Compared with BaMg$_2$Bi$_2$, CaMg$_2$Bi$_2$ exhibits a lower peak intensity in the differential $\kappa_P$ spectrum, indicating a weaker contribution from heat-carrying phonons and thereby accounting for its lower lattice thermal conductivity.
\begin{figure*}[t]
    \centering
    \includegraphics[width=1\linewidth]{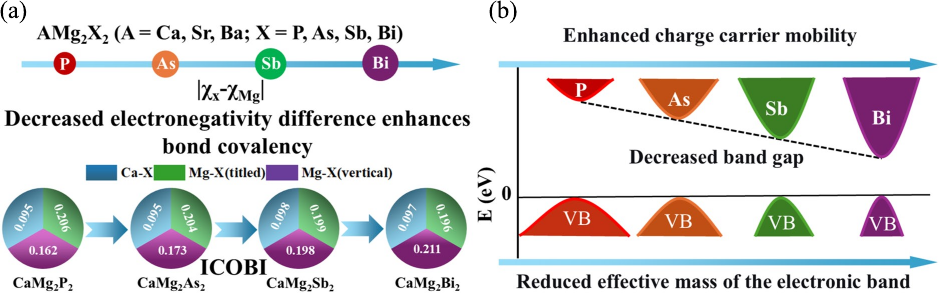}
    \caption{(a) The variation in the electronegativity difference between the X-site atom and Mg in AMg$_2$X$_2$ (A = Ca, Sr, Ba; X = P, As, Sb, Bi) and its effect on chemical bonding, where a reduced electronegativity difference enhances the covalency of the vertical Mg–X bond and leads to increased ICOBI values. (b) A schematic illustration of the band structure evolution with increasing atomic number of X, where the band gap decreases and the band dispersion is enhanced, thereby reducing the effective mass and improving carrier mobility.}
    \label{fig3}
\end{figure*}

Since $\kappa_P$ is primarily governed by phonon group velocities and lifetimes, we further projected the phonon group velocity onto each phonon mode, as shown in FIG. S12 of the Supporting Information. The projected group velocities decrease progressively with increasing X-site atomic mass, consistent with conventional mass-dependent phonon softening. In contrast, the anomalous A-site trend cannot be explained by atomic mass alone. Although Ca is lighter than Ba, the Ca-related vibrational contributions are mainly located in the mid-frequency region and therefore have a limited influence on the low-frequency phonons that dominate heat transport. Moreover, the heat-carrying low-frequency phonons in CaMg$_2$Bi$_2$ exhibit a lower cutoff frequency and reduced group velocities compared with those in BaMg$_2$Bi$_2$. These results indicate that suppressed low-frequency phonon group velocities are one of the key factors responsible for the ultralow $\kappa_L$ of CaMg$_2$Bi$_2$.

The phonon lifetime is governed by both the intrinsic scattering strength among phonons and the number of available scattering channels. The former can be characterized by anharmonicity-related descriptors, such as the anharmonicity measure $\sigma^{A}$ \cite{knoop2023anharmonicity} and mode Grüneisen parameters \cite{elalfy2021metavalent}, whereas the latter is described by the weighted phase space, which quantifies the allowed three- and four-phonon scattering channels under energy and momentum conservation \cite{li2014lattice, yuefour}. As shown in FIG.~\ref{fig2}(c), the calculated $\sigma^{A}$ values increase with temperature for all compounds, indicating progressively enhanced lattice anharmonicity at elevated temperatures. A comparison between CaMg$_2$Bi$_2$ and BaMg$_2$Bi$_2$ further reveals substantially stronger anharmonicity in CaMg$_2$Bi$_2$, which is expected to shorten phonon lifetimes and suppress particle-like thermal transport.

To further corroborate this difference in anharmonicity, the mode Grüneisen parameters were projected onto the phonon dispersions of CaMg$_2$Bi$_2$ and BaMg$_2$Bi$_2$, as shown in FIG.~\ref{fig2}(d). In the frequency range that dominates $\kappa_P$, CaMg$_2$Bi$_2$ exhibits markedly larger Grüneisen parameters than BaMg$_2$Bi$_2$, confirming its stronger lattice anharmonicity. This trend is consistent with the anharmonicity measure $\sigma^{A}$ and provides a microscopic explanation for the more strongly suppressed phonon lifetimes in CaMg$_2$Bi$_2$. 

To further clarify the origin of phonon lifetime suppression, we examined the frequency-dependent scattering rates and weighted phase spaces of all AMg$_2$X$_2$ compounds, as summarized in FIGs. S13 and S14 of the Supporting Information. A pronounced trend is observed across the series: both the four-phonon-weighted phase space and the corresponding scattering rates increase progressively with increasing X-site atomic mass. This behavior is closely associated with the enlarged phonon band gap, which modifies the available scattering channels and enhances the relative importance of four-phonon processes. 

As shown in FIG.~\ref{fig2}(e), the scattering rates of all phonon modes remain below the Ioffe–Regel limit, confirming the validity of the phonon quasiparticle picture for describing particle-like thermal transport. For both CaMg$_2$Bi$_2$ and BaMg$_2$Bi$_2$, the four-phonon scattering rates are substantially higher than the corresponding three-phonon scattering rates, especially in the low-frequency region that dominates $\kappa_P$. This indicates that four-phonon processes play a decisive role in limiting heat transport in these compounds.

The corresponding weighted phase spaces, shown in FIG.~\ref{fig2}(f), further support this conclusion. Compared with three-phonon processes, four-phonon scattering exhibits a much larger weighted phase space, reflecting the greater number of allowed scattering channels under energy and momentum conservation. Notably, in the low-frequency region, CaMg$_2$Bi$_2$ shows a slightly larger four-phonon weighted phase space than BaMg$_2$Bi$_2$. This difference can be attributed to the Ca-related vibrational modes located around 140 cm$^{-1}$, which lie within the phonon band-gap region and provide additional channels for four-phonon scattering. Together with the stronger intrinsic anharmonicity discussed above, the enlarged four-phonon phase space leads to higher total scattering rates in CaMg$_2$Bi$_2$, thereby contributing to its lower $\kappa_P$ compared with BaMg$_2$Bi$_2$.

As shown in FIG.~\ref{fig3}(a), the variation in the X-site anion has a pronounced influence on the chemical bonding characteristics of AMg$_2$X$_2$ compounds. In CaMg$_2$X$_2$, as X changes from P to As, Sb, and Bi, the electronegativity of the anion gradually decreases, accompanied by a progressive reduction in the electronegativity difference between X and Mg ($|\chi_X-\chi_{Mg}|$). This reduction in electronegativity difference consequently enhances the covalent character of the Mg–X bonds. To quantitatively characterize the bonding nature, the integrated crystal orbital bond index (ICOBI) values were calculated for the Ca–X ($d$$_1$) bond, the titled Mg–X ($d$$_2$) bond, and the vertical Mg–X ($d$$_3$) bond, where a larger ICOBI value indicates stronger covalency. As illustrated in FIG.~\ref{fig3}(a), the ICOBI values of the $d$$_1$ and $d$$_2$ bonds show only minor variations with changing anion species, fluctuating within narrow ranges of approximately 0.095–0.097 and 0.196–0.206, respectively. The small changes in ICOBI indicate that the covalent character of these two bonds is relatively insensitive to anion substitution. In comparison, the ICOBI value of the $d$$_3$ bond shows a clear monotonic increase, rising from 0.162 in CaMg$_2$P$_2$ to 0.211 in CaMg$_2$Bi$_2$, indicating a continuous strengthening of its covalent character. These results suggest that the reduction in electronegativity difference primarily reinforces the orbital hybridization associated with the $d$$_3$ bond, whereas its influence on the $d$$_1$ and $d$$_2$ bonds is comparatively limited. 
\begin{figure*}[htbp!]
    \centering
    \includegraphics[width=1\linewidth]{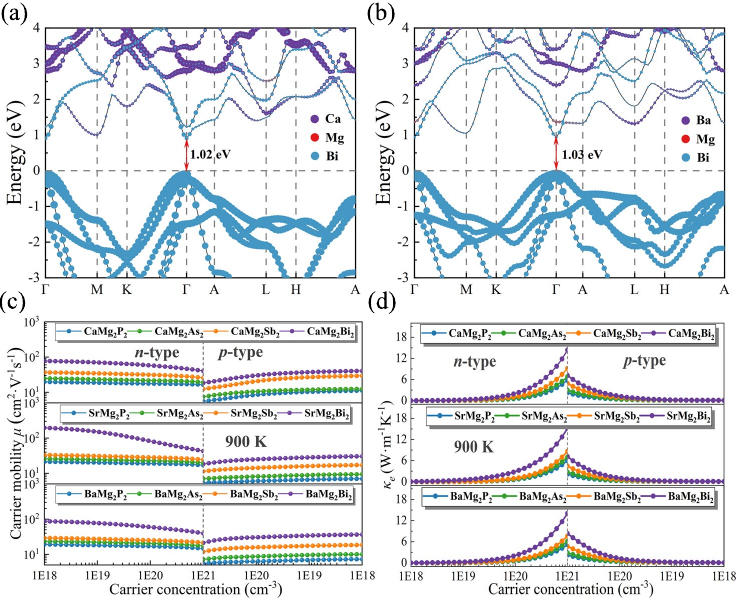}
    \caption{The atom-projected band structure of (a) CaMg$_2$Bi$_2$ and (b) BaMg$_2$Bi$_2$ calculated using the HSEsol functional. Bubble sizes represent the electronic state contributions from Ca/Ba, Mg, and Bi. (c) Carrier mobility as a function of carrier concentration at 900 K for AMg$_2$X$_2$, with the left panel showing $n$-type mobility and the right panel showing $p$-type mobility. (d) Carrier thermal conductivity as a function of carrier concentration at 900 K for AMg$_2$X$_2$, with the left panel showing $n$-type carrier thermal conductivity and the right panel showing $p$-type carrier thermal conductivity.}
    \label{fig4}
\end{figure*}

To obtain a more accurate description of the electronic structure, the band structures were further evaluated using the HSEsol functional, considering the band gap underestimation of PBEsol. As shown in FIG.~\ref{fig3}(b), the band gap exhibits a clear decreasing trend with decreasing electronegativity of the X-site element (from P to Bi). The reduced band gap facilitates carrier excitation across the bands, which is beneficial for enhancing electrical conductivity. Detailed band structures are provided in FIG. S15. More importantly, the evolution of the electronic band structure indicates a gradual increase in band dispersion near the band edges. This behavior can be attributed to the enhanced covalent character of chemical bonding and the increased orbital overlap between constituent atoms, which consequently leads to a reduction in the effective mass of charge carriers. This trend is evident in the calculated effective masses, which exhibit a pronounced decrease. Specifically, along the $\Gamma$–A direction, the electron effective mass decreases from 0.140 $m_0$ in CaMgP$_2$ to 0.050 $m_0$ in CaMg$_2$Bi$_2$, while along the $\Gamma$–K direction, it decreases from 0.932 $m_0$ to 0.503 $m_0$. Similar reductions are also observed in the SrMg$_2$X$_2$ and BaMg$_2$X$_2$ systems, as summarized in Table S1. The substantial decrease in effective mass indicates enhanced carrier mobility. Additionally, the effective masses exhibit clear anisotropy along different crystallographic directions, such as the M-$\Gamma$ and M-K directions, reflecting the intrinsically anisotropic electronic structure of the layered system. Overall, the combined effects of band gap narrowing and reduced effective mass synergistically contribute to the enhanced carrier transport performance in AMg$_2$X$_2$ compounds with heavier X-site elements.
\begin{figure*}[htbp!]
    \centering
    \includegraphics[width=1\linewidth]{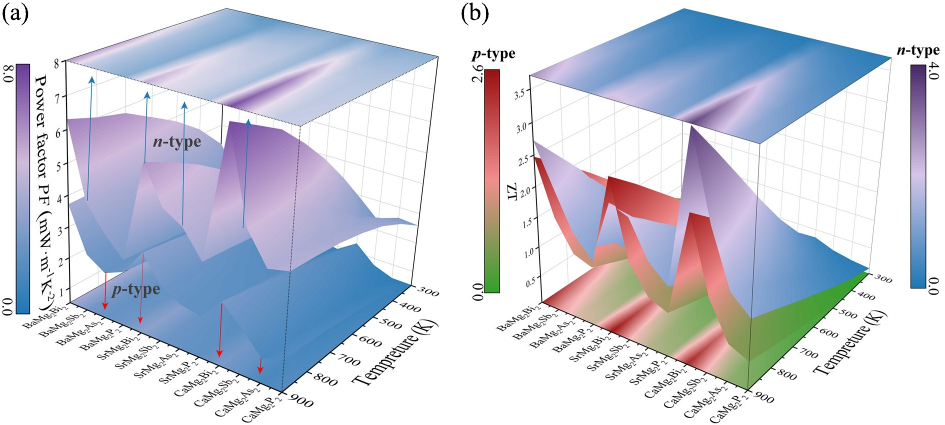}
    \caption{(a) $PF$ of AMg$_2$X$_2$ as a function of temperature (300–900 K) at the optimal carrier concentration for both $n$-type and $p$-type doping. The 3D surfaces represent $PF$ values, while the projections on the top and bottom planes correspond to the $n$-type and $p$-type $PF$ distributions, respectively. (b) $ZT$ of AMg$_2$X$_2$ under the same conditions. The 3D surfaces show $ZT$ values, with the top and bottom plane projections representing $n$-type and $p$-type $ZT$ distributions, respectively.}
    \label{fig5}
\end{figure*}

To further elucidate the electronic-structure origin of the enhanced carrier transport, the atom-projected band structures of representative CaMg$_2$Bi$_2$ and BaMg$_2$Bi$_2$ compounds are shown in FIGs.~\ref{fig4}(a) and (b), while the complete band structures of all the studied AMg$_2$X$_2$ family are summarized in FIG. S15. Overall, as the X-site anion changes from P to As, Sb, and Bi, the band gap gradually decreases. Meanwhile, the conduction-band minimum (CBM) gradually shifts toward the $\Gamma$ point. In particular, the band gaps of CaMg$_2$Bi$_2$ and BaMg$_2$Bi$_2$ decrease to 1.02 eV and 1.03 eV, respectively. In addition to the reduced band gap, CaMg$_2$Bi$_2$ and BaMg$_2$Bi$_2$ also exhibit more converged conduction bands near the Fermi level, as shown in FIGs.~\ref{fig4}(a) and (b). Such band convergence is beneficial for increasing the valley degeneracy and thereby improving the electrical transport performance under $n$-type doping~\cite{Hu2021-Pb7Bi4Se13-NC, yue2024hierarchical}. Further analysis of the atom-projected electronic states reveals that the orbital character near the band edges also changes with anion substitution. For CaMg$_2$P$_2$, CaMg$_2$As$_2$, and CaMg$_2$Sb$_2$, the CBM near the M point mainly originates from hybridized contributions of Ca and X atoms. As a comparison, both the CBM and VBM near the $\Gamma$ point are predominantly contributed by Bi states for CaMg$_2$Bi$_2$, as implied in FIG.~\ref{fig4}(a). Similar orbital characteristics are also observed in BaMg$_2$Bi$_2$ (FIG.~\ref{fig4}(b)). These results indicate that heavier anion substitution significantly modifies the electronic structure near the band edges, thereby affecting the carrier transport behavior in the AMg$_2$X$_2$ compounds.
 
Based on the calculated electronic band structures, we further evaluated the carrier mobility of AMg$_2$X$_2$ compounds at 900 K as a function of carrier concentration, as shown in FIG.~\ref{fig4}(c). Overall, the $n$-type mobility is higher than the $p$-type mobility for all compounds, indicating more favorable electron transport in this system. For a fixed A-site cation, the carrier mobility generally increases as the X-site anion changes from P to Bi, which is consistent with the enhanced band dispersion and reduced carrier effective mass discussed above. In addition, the mobility gradually decreases with increasing carrier concentration, mainly due to the strengthened carrier scattering at high doping levels. The corresponding carrier thermal conductivity $\kappa_e$ is shown in FIG.~\ref{fig4}(d). Different from the evolution of mobility, $\kappa_e$ increases significantly with carrier concentration, especially near the high-concentration region. According to the Wiedemann–Franz relation, $\kappa_e=L\sigma T$, the electronic thermal conductivity is directly proportional to the electrical conductivity at a given temperature \cite{Franz}. Therefore, the enhanced carrier concentration leads to an increase in $\sigma$, which consequently results in the increase of $\kappa_e$. Therefore, although the improved carrier mobility is beneficial for achieving a high power factor, it also leads to an increased electronic thermal conductivity, which may partially offset the positive contribution of enhanced electrical transport to $ZT$. 

\begin{figure*}[htbp!]
    \centering
    \includegraphics[width=1\linewidth]{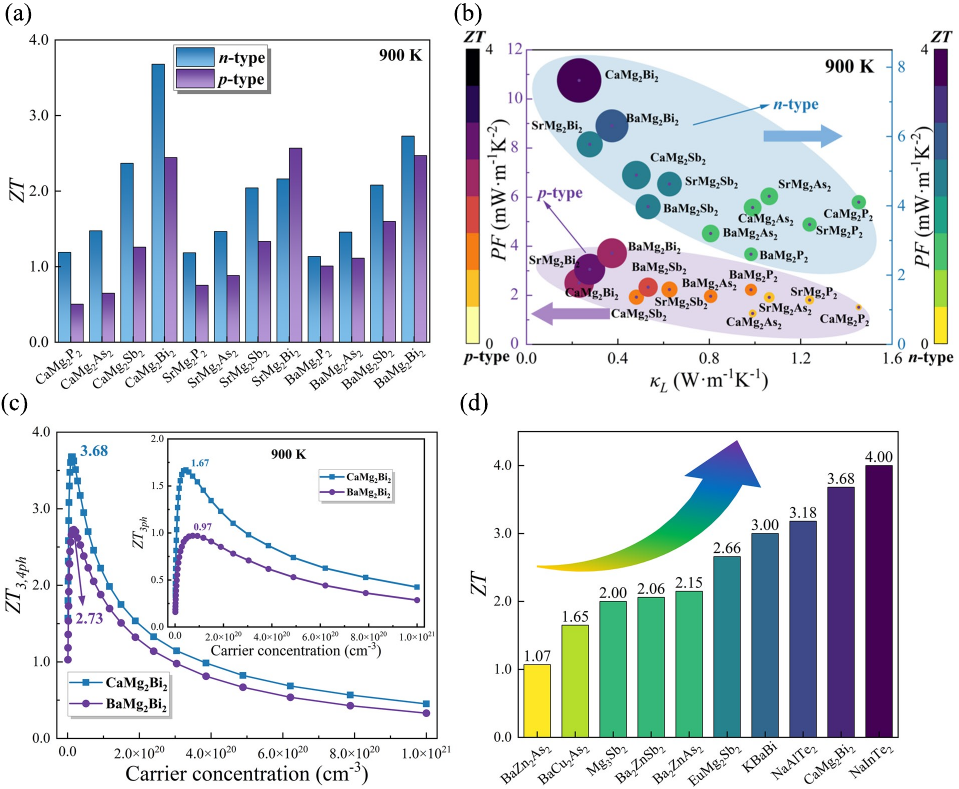}
    \caption{(a) Maximum $ZT$ of AMg$_2$X$_2$ compounds at 900 K. The blue and purple bars represent the optimized $n$-type and $p$-type $ZT$ values, respectively. (b) Relationship between the $PF$ and $\kappa_L$ of AMg$_2$X$_2$ compounds at 900 K. The bubble size represents the magnitude of $ZT$, while the cyan and purple shaded regions denote the optimized $n$-type and $p$-type $ZT$ values, respectively. (c) Carrier concentration dependence of the calculated $ZT_{3,4ph}$ values for CaMg$_2$Bi$_2$ and BaMg$_2$Bi$_2$ at 900 K, where $ZT_{3,4ph}$ represents the results obtained by considering both three-phonon and four-phonon scattering. The inset shows the corresponding $ZT_{3ph}$ results obtained by considering only three-phonon scattering. (d) Comparison of the peak $ZT$ value of CaMg$_2$Bi$_2$ with those reported for representative Zintl thermoelectric compounds at high temperatures, where the specific compounds corresponding to the labels are detailed in Refs \cite{zhang2017influence, shao2025gd, zhang2025diffusons, song2025band, feng2021zintl, yao2026ultralow, szeleszczuk2026computational, wang2024high}. }
    \label{fig6}
\end{figure*}

After accounting for the dominant scattering mechanisms, including acoustic deformation potential, polar optical phonon, and ionized impurity scattering, the $PF$ of AMg$_2$X$_2$ compounds were evaluated, as shown in FIG.~\ref{fig5}(a). Overall, the $PF$ under $n$-type doping is significantly higher than that under $p$-type doping, indicating more favorable electron transport in this system. This behavior can be attributed to the combined effects of enhanced carrier mobility and band convergence near the conduction band edge. Correspondingly, for a fixed A-site cation, the $PF$ exhibits a clear dependence on X-site anion substitution, increasing progressively as the X-site anion varies from P to Bi. This originates from the evolution of the electronic structure. The enhanced band convergence near the conduction band edge increases the valley degeneracy and is beneficial for a larger Seebeck coefficient ($S$). Furthermore, the increased band dispersion near the band edges and reduced band gap is favorable for electronic transport, leading high electrical conductivity ($\sigma$). As a result, the electronic structure becomes more conducive to achieving a favorable balance between $\sigma$ and $S$, leading to enhanced $PF$. In addition, the $PF$ exhibits an overall increasing trend with temperature, with a noticeable enhancement at elevated temperatures. Notably, at 900 K, CaMg$_2$Bi$_2$ achieves a maximum $n$-type $PF$ of $7.62~\mathrm{mW\cdot\,m^{-1}\,K^{-2}}$ at a carrier concentration of $7.88\times10^{20}$ cm$^{-3}$, demonstrating excellent electronic transport performance.

Ultimately, by combining the thermal and electrical transport properties, we obtained the $ZT$ values of all the investigated materials. As shown in FIG.~\ref{fig5}(b), the $ZT$ values of AMg$_2$X$_2$ compounds increase significantly with temperature, indicating their promising thermoelectric performance at elevated temperatures. In most cases, the $n$-type compounds exhibit higher $ZT$ values than their $p$-type counterparts, which is consistent with their superior power factors. However, SrMg$_2$Bi$_2$ shows exceptional behavior, where the optimized $p$-type $ZT$ exceeds the corresponding $n$-type value. This is mainly because the $p$-type SrMg$_2$Bi$_2$ exhibits a lower carrier thermal conductivity $\kappa_e$ than the $n$-type case, which reduces the total thermal conductivity and thus leads to a higher optimized $ZT$. Among all the investigated compounds, CaMg$_2$Bi$_2$ shows the most promising $n$-type thermoelectric performance at high temperature, which originates from the synergistic effect of enhanced electrical transport and ultralow lattice thermal conductivity.

FIG.~\ref{fig6}(a) presents the maximum $ZT$ values of all compounds at 900 K. The optimized $ZT$ of AMg$_2$X$_2$ compounds exhibit a pronounced dependence on the X-site anion. For a fixed A-site cation, the $ZT$ values show a progressive enhancement as the X-site anion varies from P to Bi. Among all the investigated compounds, CaMg$_2$Bi$_2$ achieves the highest optimized $n$-type $ZT$ value of 3.68 at 900 K, indicating its great potential for thermoelectric applications. To further reveal the origin of the high $ZT$ values, FIG.~\ref{fig6}(b) presents the relationship between the $PF$ and $\kappa_L$ at 900 K. The $n$-type compounds are generally located in the high power factor region, whereas the $p$-type compounds exhibit relatively lower power factors. For $n$-type CaMg$_2$Bi$_2$, the coexistence of a high power factor and an ultralow lattice thermal conductivity enables the best thermoelectric performance among the AMg$_2$X$_2$ compounds.

FIG.~\ref{fig6}(c) further compares the carrier concentration dependent $ZT$ values of CaMg$_2$Bi$_2$ and BaMg$_2$Bi$_2$ at 900 K. When both three-phonon and four-phonon scattering processes are included, CaMg$_2$Bi$_2$ reaches a peak $ZT$ of 3.68 at a carrier concentration of 1.37 $\times$ 10$^{19}$ cm$^{-3}$, while BaMg$_2$Bi$_2$ achieves a peak $ZT$ of 2.73 at a carrier concentration of 1.74 $\times$ 10$^{19}$ cm$^{-3}$. In contrast, when only three-phonon scattering is considered, the corresponding peak $ZT$ values are only 1.67 and 0.97, respectively, as shown in the inset. These results indicate that four-phonon scattering is essential for accurately evaluating the lattice thermal conductivity and thermoelectric performance of strongly anharmonic AMg$_2$X$_2$ compounds. Finally, FIG.~\ref{fig6}(d) compares the peak $ZT$ value of CaMg$_2$Bi$_2$ with those of representative Zintl thermoelectric compounds. The predicted $ZT$ value of CaMg$_2$Bi$_2$ is higher than many reported Zintl compounds. These results further confirm that AMg$_2$X$_2$ compounds, especially CaMg$_2$Bi$_2$, are competitive candidates for thermoelectric applications.

In summary, we systematically investigated the thermoelectric transport properties of CaAl$_2$Si$_2$-type AMg$_2$X$_2$ (A = Ca, Sr, Ba; X = P, As, Sb, Bi) Zintl compounds by combining first-principles calculations, machine-learning interatomic potentials and Boltzmann transport theory. The results show that anion engineering significantly affects the thermal transport behavior of AMg$_2$X$_2$ compounds by regulating chemical bonding and phonon dispersion. As the X-site anion changes from P to Bi, the enlarged phonon band gap restricts the available three-phonon scattering channels and enhances the relative contribution of four-phonon scattering, leading to strongly suppressed $\kappa_L$. As a result, CaMg$_2$Bi$_2$ achieves an ultralow $\kappa_L$ of $0.23~\mathrm{W\cdot\,m^{-1}\,K^{-1}}$ at 900 K. In addition to suppressing phonon transport, anion engineering also optimizes the electronic transport properties. As the X-site anion changes from P to Bi, the reduced band gap and enhanced band dispersion facilitate carrier transport, while the increased conduction band convergence is beneficial for maintaining a large Seebeck coefficient. This anion-driven optimization of the electronic band structure results in a progressive enhancement of the $PF$, with CaMg$_2$Bi$_2$ achieving the highest $n$-type $PF$ of $7.62~\mathrm{mW\cdot\,m^{-1}\,K^{-2}}$ at 900 K. Anion-driven concurrent optimization of phonon and electronic transport enables CaMg$_2$Bi$_2$ to achieve a maximum $n$-type $ZT$ of 3.68 at 900 K. These findings identify CaMg$_2$Bi$_2$ as a promising thermoelectric candidate and demonstrate that anion engineering is an effective strategy for optimizing the thermoelectric performance of CaAl$_2$Si$_2$-type Zintl compounds.

\section{Acknowledgement}
N.W. and Y.L. acknowledge the National Key Research and Development Program of China (No. 2023YFA1406200) and the Institute of High-pressure Physics of Ningbo University for its computational resources. The work carried out in Spain forms part of the national research project PID2025-173011NA-I00, funded by MICIU/AEI/10.13039/501100011033. J.Y. acknowledges the China Scholarship Council (No. 202408330112). Y.-W.F. acknowledges the financial support received from the IKUR Strategy under the collaboration agreement between Ikerbasque Foundation and Centro de Física de Materiales (CFM-MPC) on behalf of the Department of Science, Universities and Innovation of the Basque Government (HPCAI21: AI-CrysPred). Y.-W.F. is partially supported by the Extraordinary Grant of CSIC (No. 2025ICT122). W.R. gratefully acknowledges the computing time provided by the NHR Center NHR4CES at RWTH Aachen University under project p0027528, with funding from the German Federal Ministry of Research, Technology and Space and the German federal states through the National High-Performance Computing (NHR) joint funding program. S.L. acknowledges the computation resources provided by the National Academic Infrastructure for Supercomputing in Sweden (NAISS) -- at supercomputer centers NSC and PDC, partially funded by the Swedish Research Council through grant agreement No. 2022-06725 -- and by the Vienna Scientific Cluster (VSC) in Austria.

\section{Conflict of Interest}
The authors declare no conflict of interest.

\section{Author Contributions}
N.W. and J.Y. contributed equally to this work. N.W. and J.Y. conceived the project. H.Z., W.R., S.L. and X.S. tested and verified the data. N.W., J.Y., and Y.-W.F. co-wrote the manuscript. Y.-W.F., Y.L., and T.C. supervised the research and acquired funding. All authors discussed the results and commented on the manuscript.

	\nocite{*}
	\bibliography{cas-refs}
	
\end{document}